\documentclass[twocolumn,a4paper,showpacs,preprintnumbers,prc,%
amsmath,amssymb,byrevtex]{revtex4}
\begin{document}
\title{Radiative capture of polarized neutrons by polarized protons}
\author{G. Ramachandran}
\affiliation{Indian Institute of Astrophysics, Koramangala, Bangalore-560 034, India}
\author{P. N. Deepak}
\email{pndeepak@sancharnet.in}
\affiliation{Department of Studies in Physics, University of Mysore,
Mysore-570 006, India}
\begin{abstract}
A model-independent irreducible tensor approach to $\vec{p}(\vec{n},\gamma)d$
is presented and an explicit form for the spin-structure of the matrix 
$\boldsymbol{M}$ for the reaction is obtained in terms of the Pauli spin-matrices 
$\boldsymbol{\sigma}(n)$ and $\boldsymbol{\sigma}(p)$.  Expressing the multipole 
amplitudes in terms of the triplet $\to$ triplet and singlet $\to$ triplet
transitions, we point out how the initial singlet and triplet contributions
to the differential cross section can be determined empirically.  
\end{abstract}
\maketitle

The study of radiative capture of neutrons by protons is of interest not 
only to nuclear physics as a testing ground for theories of $NN$ interaction, 
but also to astrophysics, where the fusion reaction is part of the proton-proton 
chain responsible for the generation of solar energy and production of 
elements in the early universe \cite{astro}.  The cross section for the 
process was directly measured \cite{cs} for the first time at neutron 
energy of 550 keV, although measurements \cite{cs1} with thermal neutrons 
have been carried out earlier.  The 10\% discrepancy noted between such 
measurements and theory was accounted for with surprising accuracy by 
the inclusion \cite{mec} of meson exchange currents (MEC).  Measurements 
\cite{cs2} at higher energies were used to test effects of $\rho$ and 
$\omega$ exchanges and relativistic corrections to the impulse approximation
\cite{ia}.  Although a thorough review of the inverse reaction $d(\gamma,n)p$ 
over a wide range of energies is found in \cite{inv}, the energy region 
just above threshold does not seem to have received much attention.  
Photodisintegration experiments at low energies focused attention 
\cite{inv-expts} on the relative $M1$ and $E1$ contributions.  Experiments 
employing polarized photons have been reported between 5 to 10 MeV \cite{phot}
and more recently \cite{phot-rec} at 3.58 MeV.  The angular distribution 
as well as polarization of the neutron in $d(\gamma,p)n$ were measured 
\cite{angular}.  Though the measured angular distribution of the neutron 
was found \cite{ang-neu} to be in good agreement with theory at 2.75 MeV, the 
angular distribution and neutron polarization in \cite{angular} differ 
from theory which includes the MEC contributions.  Employing polarized 
neutrons at 6 MeV and 13.4 MeV, the analyzing powers \cite{ang-neu}
in $p(\vec{n},\gamma)d$ was measured which differed from theory.  Recent 
cross section calculations \cite{recent1,recent2}, which agree with 
each other within 5\% deviation, are found to differ from the 1967 
estimates of Fowler et al \cite{astro}.  The cross section was obtained 
in \cite{recent1} by fitting the existing data with a polynomial expansion, 
while the calculation in \cite{recent2} includes MEC's, isobar currents 
and pair currents.  The theory is in good agreement with cross section 
data for neutrons with energy above 14 MeV, but deviates by about 
15\% from older $d(\gamma,p)n$ measurements between 2.5 MeV and 2.75 MeV 
\cite{old}, which correspond respectively to neutron energies 550 keV 
and 1080 keV in $p(n,\gamma)d$.  Apart from the fact that uncertainties 
in $p(n,\gamma)d$ cross sections at energies below about 600 keV lead
to dominant uncertainties in the determination of the relative abundances 
of elements in the early universe, it is interesting to study the fusion 
reaction for its own sake, especially since a beginning has also been made 
\cite{cs} to study $p(n,\gamma)d$ precisely in the region of energies of 
interest to astrophysics.  This recent study \cite{cs} has shown 
experimentally that the $M1$ transition from S-wave capture, which dominates 
at lower energies and the $E1$ transition from P-wave capture at higher 
energies are comparable at 550 keV.

\begin{table}
\caption{Transitions from initial states of $np$ system, from 
threshold onwards, to the final $^2H$ state with spin-parity 
$1^+$ and isospin $I=0$, together with the corresponding
$\Delta I,\ \Delta s$ and multipole characteristics of the 
radiation emitted in the fusion reaction.}
\label{tab-np-1}
\begin{ruledtabular}
\begin{tabular}{ccccccl}
Initial state && $\Delta I$ && $\Delta s$ && Allowed multipolarities\\
&&&&&& of emitted radiation\\[.2cm]\hline\hline\\
$^1S_0\quad I=1$ && $1$ && $1$ && $M1$\\
$^3S_1\quad I=0$ && $0$ && $0$ && $M1\quad E2$\\
$^1P_1\quad I=0$ && $0$ && $1$ && $E1\quad M2$\\
$^3P_0\quad I=1$ && $1$ && $0$ && $E1$\\
$^3P_1\quad I=1$ && $1$ && $0$ && $E1\quad M2$\\
$^3P_2\quad I=1$ && $1$ && $0$ && $E1\quad M2\quad E3$\\
\end{tabular}
\end{ruledtabular}\\[.15cm]
\hspace*{4cm} ... and so on.
\end{table}
It is interesting to observe that the neutron capture in $p(n,\gamma)d$ 
from the $^{3}S_1$ state is from the isospin $I=0$ state, which is the 
same for the deuteron, while that from the $^{1}S_0,\ I=1$ state is 
characterized by a $\Delta I=1$ transition.  Likewise, capture from 
$^{1}P_1,\ I=0$ state leads to a $\Delta I=0$ transition, whereas those from
$^{3}P_{j=0,1,2},\ I=1$ states are characterized by $\Delta I=1$ transitions 
(see TABLE \ref{tab-np-1}).  The cross section at very low energies is 
dominated by the $\Delta I=1$ amplitude for the $M1$ transition from 
the $^1S_0$ state, since the $NN$ scattering length in the $^1S_0$ channel 
is large and the $M1$ amplitude is proportional to the isovector magnetic 
moment of the nucleon, which is more than five times larger than the 
isoscalar magnetic moment.  It was, however, noted quite early 
\cite{breit-rustgi} that the spin-dependent effects in the fusion reaction 
are sensitive to the small $\Delta I=0$ amplitudes $M1$ and $E2$.  
Hence a ``polarized-target-beam test'' was proposed for the $^3S_1\to ^3S_1$ 
radiative transitions in thermal $np$ capture.  The isoscalar $M1$ 
and $E2$ amplitudes have also been studied theoretically recently 
\cite{J-W}, using effective field theory, since the circular polarization
\cite{bazhanov} of photons emitted in the capture of polarized neutrons 
by unpolarized protons and the angular distribution \cite{muller} in the 
capture of polarized neutrons by polarized protons are sensitive to 
the presence of these amplitudes.  Parameter-free predictions, employing 
the Weinberg \cite{wein} scheme of power counting, have also been made 
\cite{park} for the spin-dependent observables in $\vec{p}(\vec{n},\gamma)d$ 
at threshold energies.  However, the recently \cite{muller} measured 
value of $\eta=(1.0\pm\,2.5)\times 10^{-4}$ for $\gamma$-anisotropy at
$50.5\%$ polarization of neutrons and protons in $\vec{p}(\vec{n},\gamma)d$ 
cannot test the effective field theory predictions at $10^{-7}$.  
Although this measurement provides the first experimental value for a 
spin-dependent observable in $\vec{p}(\vec{n},\gamma)d$, spin-observables 
in elastic $\vec{N}\vec{N}$ scattering have been measured for more
than two decades \cite{bug}.  More recently, attention was focused on such
observables in $\vec{N}\vec{N}$ scattering, with a veiw to determine the exact
strength of the tensor interaction and the case of $\pi^\circ$ production in 
$\vec{p}\vec{p}$ collisions \cite{pp} for which the threshold itself is high. 
Even at higher energies, it is clear that the singlet and triplet radiative 
captures from any arbitrary initial partial wave, $\ell$ lead to $p(n,\gamma)d$ 
characterized respectively by $\Delta I=0$ and $\Delta I=1$ or $\Delta I=1$ 
and $\Delta I=0$ depending on whether the initial parity $(-1)^\ell$ is odd or even.
Since the $NN$ interaction, as elicited from elastic scattering data, conserves 
channel spin, $s$ (apart from total angular momentum $j$) as well as isospin 
$I$, it would be of interest to study experimentally the relative strengths of 
the initial singlet and triplet contributions to the $p(n,\gamma)d$
reaction at any given energy, since the fusion reaction leads to transitions 
in the two-nucleon system, which change spin $s$ as well as isospin $I$.  
It is also worth noting that $np$ fusion near threshold bears some resemblance 
to the case of $pd$ fusion, where the validity of the so-called 
``no-quartet theorem'' \cite{cjr} was questioned and led to several 
incisive theoretical studies \cite{friar} in later years.  More recently, 
it was pointed out \cite{grpndsp} that a non-zero tensor analyzing power in 
$\overrightarrow{p\phantom{i}}(\overrightarrow{d\ },\gamma)\,^3He$ 
by itself provides a clear signature to the 
contributions from the quartet amplitudes.  Moreover, it was shown 
\cite{grpnd} that it is possible to determine empirically the individual 
doublet and quartet differential cross sections at any given energy through 
appropriate measurements of the relevant spin-observables in $\vec{N}\vec{d}$ 
fusion.  In contrast to the difficult $\vec{N}\vec{d}$ fusion experiments 
suggested in \cite{grpnd}, it should be technically more feasible to carry out 
the $\vec{p}(\vec{n},\gamma)\,^2H$ experiments \cite{muller}.      

The purpose of this Rapid Communication is to present a model-independent 
theoretical approach to the $np$ fusion process based on the irreducible 
tensor formalism and identify observables in $\vec{p}(\vec{n},d)\gamma$ 
experiments that facilitate determination of the singlet and triplet cross 
sections empirically for the fusion reaction at any energy. 

Let $\boldsymbol{p}(p,0,0)$ and $\boldsymbol{k}(k,\theta,0)$ denote 
respectively the neutron and photon momenta in a right-handed c.m. frame 
\cite{mad} and let $\langle m_d;\boldsymbol{k}\mu|\boldsymbol{T}|
\boldsymbol{p};m_n,m_p\rangle$ denote the elements of the on-energy-shell 
$\boldsymbol{T}$-matrix for the fusion reaction $p(n,\gamma)d$ where $m_n,m_p,m_d$ denote 
respectively the spin-projections of the neutron, proton and deuteron and 
$\mu=\pm 1$ denote the left- and right-circular states of photon polarization, 
as defined by Rose \cite{rose}.  The unpolarized differential cross section 
for the reaction is then given by 
\begin{equation}
\label{udcs}
\begin{split}
\frac{\textrm{d}\sigma_0}{\textrm{d}\Omega}=&\left(\frac{k}{2\pi E}\right)^2
\ \frac{E_n E_p E_d}{p}\ \frac{1}{4}\sum_{m_n,m_p=-\frac{1}{2}}^{\frac{1}{2}}
\ \sum_{m_d=-1}^{1}\\&\times\sum_{\mu=-1,1}
\ |\langle m_d;\boldsymbol{k}\mu|\boldsymbol{T}|\boldsymbol{p};m_n,m_p\rangle|^2,
\end{split}
\end{equation}    
where $E_p,E_n$ and $E_d$ denote the c.m. energies of the proton, neutron 
and $^2H$ respectively, when the reaction takes place at c.m. energy $E$.  
Expressing the initial $np$ system in terms of isosinglet and isotriplet 
states and making use of the standard multipole expansion \cite{rose} for 
the photon in the final state and the usual partial wave expansion for 
relative motion in the initial state with channel-spin $s=0,1$, the 
$\boldsymbol{T}$-matrix may be expressed as 
\begin{equation}
\label{T}
\boldsymbol{T}=\sum_{s=0,1}\ \sum_{\lambda=|1-s|}^{(1+s)}
\ \sum_{\mu=-1,1} \left(S^\lambda(1,s)\cdot T^\lambda(\mu,s)
\right),  
\end{equation}
where $S^\lambda_\nu(1,s)$ denote \cite{gr-msv} irreducible tensor operators 
of rank $\lambda$ in hadron spin-space.  The irreducible tensor amplitudes 
$T^\lambda_\nu(\mu,s)$ in Eq. \eqref{T} are given explicitly by
\begin{align}
\label{partial-wave}
T^\lambda_\nu&(\mu,s)=\sqrt{\frac{2}{3}}\,(2\pi)(-1)^{1-s}
\sum_{\ell=0}^{\infty}\ \sum_{L=1}^{\infty}
\ \sum_{j=|\ell-s|}^{(\ell+s)}\ (\textrm{i})^{\ell-L}\nonumber\\
&\times(-1)^{\lambda-L}\,[\ell]\,[L]\,[j]^2\,d^{\,L}_{\nu\mu}(\theta)
\,C(\ell L\lambda;0\nu\nu)\\&\times 
W(\ell sL1;j\lambda)\,\left[T^{j\,(\text{mag})}_{L;\ell s}(E)-\textrm{i}\mu\ T^{j\,(\text{elec})}_{L;\ell s}(E)
\right]\nonumber,
\end{align}        
where the energy dependence is carried entirely by the partial wave 
magnetic and electric $2^L$-multipole amplitudes $T^{j\,(\text{mag})}_{L;\ell s}(E)$ 
and $T^{j\,(\text{elec})}_{L;\ell s}(E)$ respectively, while the angular dependence 
is contained entirely in $d^{\,L}_{\nu\mu}(\theta)$. We use the shorthand $[j]=
{\sqrt{(2j+1)}}$ and the rest of the notations follow \cite{rose}.  The 
multipole amplitudes are given in terms of the reduced matrix elements by
\begin{equation}
\begin{split}
T^{j\,(\text{mag})}_{L;\ell s}(E)&=\frac{1}{2}\left\{(-1)^{L+1}+(-1)^\ell\right\}
\,\langle\ ||\boldsymbol{T}\,||\ \rangle\\[.2cm]
\,T^{j\,(\text{elec})}_{L;\ell s}(E)&=\frac{1}{2}\left\{(-1)^{L}+(-1)^\ell\right\}
\, \langle\ ||\boldsymbol{T}\,||\ \rangle\,,
\end{split}
\end{equation} 
where $\langle\ ||\boldsymbol{T}\,||\ \rangle$ denotes
\begin{equation}
\begin{split}
\langle\ &||\boldsymbol{T}\,||\ \rangle=\sum_{I=0,1}\,(-1)^I\,[I]^{-\frac{1}{2}}
\,\left[1-(-1)^{\ell+s+I}\right]\\
&\times
C({\textstyle{\frac{1}{2}}}
{\textstyle{\frac{1}{2}}}I;-{\textstyle{\frac{1}{2}}}{\textstyle{\frac{1}{2}}}0)
\Bigl\langle(1L)j;I_d=0\bigl|\bigl|\boldsymbol{T}\,\bigr|\bigr|(\ell s)
j;I\Bigr\rangle\,.
\end{split}
\end{equation}
Expressing the irreducible tensor operators $S^\lambda_\nu(1,s)$ of rank
$\lambda$ in terms of the Pauli spin-matrices $\boldsymbol{\sigma}(n)$
and $\boldsymbol{\sigma}(p)$ and unit $2\times 2$ matrices $\sigma^0_0(n),
\, \sigma^0_0(p)$ for neutron, proton respectively following \cite{gr-msv}, 
we may rewrite Eq. \eqref{udcs} in the elegant form
\begin{align}
\label{eleg-udcs}
\frac{\textrm{d}\sigma_0}{\textrm{d}\Omega}=\frac{1}{4}
\ \text{Tr}\left(\boldsymbol{M}\boldsymbol{M}^\dagger
\right),
\end{align}
where $\text{Tr}(\equiv \sum_{m_d}\sum_\mu)$ denotes the Trace 
or Spur and the matrix $\boldsymbol{M}$ for $p(n,\gamma)d$ has the form
\begin{widetext}
\begin{equation}
\label{sigma}
\boldsymbol{M}=\sum_{\lambda_1,\lambda_2=0}^1
\ \sum_{\lambda=|\lambda_1-\lambda_2|}
^{(\lambda_1+\lambda_2)}\ \left(
\left(\sigma^{\lambda_1}(n)\otimes\sigma^{\lambda_2}(p)\right)^\lambda\cdot
M^\lambda(\lambda_1,\lambda_2;\mu)\right),
\end{equation}
\end{widetext}
in terms of irreducible tensor amplitudes $M^\lambda_\nu(\lambda_1,\lambda_2;\mu)$
of rank $\lambda$ given by
\begin{equation}
\label{M-T}
\begin{split}
M^\lambda_\nu(\lambda_1,\lambda_2;\mu)=&\left(\frac{k}{2\pi E}\right)
\left(\frac{E_n E_p E_d}{p}\right)^{\frac{1}{2}}\,\frac{3}{2}\,[\lambda_1]
[\lambda_2]
\\&\times\sum_{s=0}^1\ [s]
\,\left\{
\begin{matrix}
\textstyle{\frac{1}{2}}&\textstyle{\frac{1}{2}}&1\\[.2cm]
\textstyle{\frac{1}{2}}&\textstyle{\frac{1}{2}}&s\\[.2cm]
\lambda_1&\lambda_2&\lambda
\end{matrix}
\right\}\,T^\lambda_\nu(\mu,s),
\end{split}
\end{equation}
where $\{\ \}$ denote Wigner-$9j$ symbols \cite{rus}.  We may
explicit Eq. \eqref{sigma} as
\begin{equation}
\label{wolf}
\begin{split}
\boldsymbol{M}=&A+B\left(\boldsymbol{\sigma}(n)\cdot\boldsymbol{\sigma}(p)\right)+
\left(\boldsymbol{\sigma}(n)+\boldsymbol{\sigma}(p)\right)\cdot\boldsymbol{C}\\&+
\left(\boldsymbol{\sigma}(n)-\boldsymbol{\sigma}(p)\right)\cdot\boldsymbol{D}+
\left(\boldsymbol{\sigma}(n)\times\boldsymbol{\sigma}(p)\right)\cdot\boldsymbol{E}\\&+
\left(\left(\boldsymbol{\sigma}(n)\otimes\boldsymbol{\sigma}(p)\right)^2\cdot F^2\right),
\end{split}
\end{equation}
where the coefficients are related to \eqref{M-T} through
\begin{equation}
\label{wolfenstein}
\begin{split}
A&=M^0_0(0,0;\mu)\ ;\ B=-\frac{1}{\sqrt{3}}M^0_0(1,1;\mu)\ ;\\[.2cm]
C^1_\nu&=\frac{1}{2}\left[M^1_\nu(1,0;\mu)+M^1_\nu(0,1;\mu)\right];\\[.2cm]
D^1_\nu&=\frac{1}{2}\left[M^1_\nu(1,0;\mu)-M^1_\nu(0,1;\mu)\right];\\[.2cm]
E^1_\nu&=\frac{\textrm{i}}{\sqrt{2}}M^1_\nu(1,1;\mu)\ ;\ F^2_\nu=M^2_\nu(1,1;\mu)
\end{split}
\end{equation} 
and $C^1_\nu,\ D^1_\nu,\ E^1_\nu$ denote respectively the spherical components 
of $\boldsymbol{C},\ \boldsymbol{D},\ \boldsymbol{E}$.  Comparing Eq. \eqref{wolf} 
with $\boldsymbol{M}$ for elastic $NN$ scattering \cite{wolf} shows clearly 
that the fourth and fifth terms containing $\left(\boldsymbol{\sigma}(n)-
\boldsymbol{\sigma}(p)\right)$ and $\left(\boldsymbol{\sigma}(n)\times
\boldsymbol{\sigma}(p)\right)$ are the ones which induce transitions from the 
initial spin-singlet state to the final spin-triplet state of the deuteron.  
To estimate the singlet and triplet cross sections empirically we observe that 
the differential cross section for $\vec{p}(\vec{n},\gamma)d$ is given by
\begin{equation}
\label{dcs}
\frac{\textrm{d}\sigma}{\textrm{d}\Omega}=\text{Tr}\left(\boldsymbol{M}\rho
\boldsymbol{M}^\dagger\right),
\end{equation}
where the density matrix
\begin{equation}
\label{dm}
\rho=\frac{1}{4}\left[1+\Bigl(\boldsymbol{\sigma}(n)\cdot\boldsymbol{P}(n)\Bigr)\right]
\ \left[1+\Bigl(\boldsymbol{\sigma}(p)\cdot\boldsymbol{P}(p)\Bigr)\right]
\end{equation}
describes the initial spin state if $\boldsymbol{P}(n)$ and $\boldsymbol{P}(p)$
denote respectively the neutron and proton polarizations.  Rewriting Eq. \eqref{dcs} as
\begin{equation}
\frac{\textrm{d}\sigma}{\textrm{d}\Omega}=\frac{1}{4}\sum_{\alpha,\beta=0,x,y,z}
\ P_{\alpha}(n)\,P_{\beta}(p)\,B_{\alpha\beta},
\end{equation} 
where $P_0(n)=1,P_0(p)=1$ and
\begin{equation}
B_{\alpha\beta}=\text{Tr}\left(\boldsymbol{M}\sigma_\alpha(n)\sigma_\beta(p)
\boldsymbol{M}^\dagger\right),
\end{equation}
we readily see that the unpolarized differential cross section \eqref{eleg-udcs}
is given by $(1/4)B_{00}$.  Noting \cite{prc} further that
\begin{align}
&\hspace*{-7.5pt}\boldsymbol{\pi}(1,1)=\frac{1}{4}\left[1+\sigma_z(n)+\sigma_z(p)+
\sigma_z(n)\sigma_z(p)
\right]\\[.2cm] 
&\hspace*{-7.5pt}\boldsymbol{\pi}(1,0)=\frac{1}{4}\left[1+\sigma_x(n)\sigma_x(p)+
\sigma_y(n)\sigma_y(p)-\sigma_z(n)\sigma_z(p)
\right]\\[.2cm]
&\hspace*{-7.5pt}\boldsymbol{\pi}(1,-1)=\frac{1}{4}\left[1-\sigma _z(n)-\sigma _z(p)
+\sigma _z(n)\sigma _z(p)
\right]\\[.2cm]
&\hspace*{-7.5pt}\boldsymbol{\pi}(0,0)=\frac{1}{4}\left[1-\sigma_x(n)\sigma_x(p)
-\sigma_y(n)\sigma_y(p)-\sigma_z(n)\sigma_z(p)
\right]
\end{align}
are the projection operators $\vert sm\rangle\langle sm\vert,\,s=0,1;\,m=
+s,\ldots,-s$, we readily identify
\begin{align}
\label{cdc1}
\frac{\textrm{d}\sigma_{1,1}}{\textrm{d}\Omega}&=\frac{1}{16}\left[B_{00}+B_{z0}
+B_{0z}+B_{zz}\right]\\[.2cm]
\label{cdc2}
\frac{\textrm{d}\sigma_{1,0}}{\textrm{d}\Omega}&=\frac{1}{16}
\left[B_{00}+B_{xx}+B_{yy}-B_{zz}\right]\\[.2cm]
\label{cdc3}
\frac{\textrm{d}\sigma_{1,-1}}{\textrm{d}\Omega}&=\frac{1}{16}
\left[B_{00}-B_{z0}-B_{0z}+B_{zz}\right]\\[.2cm]
\label{cdc4}
\frac{\textrm{d}\sigma_{0,0}}{\textrm{d}\Omega}&=\frac{1}{16}
\left[B_{00}-B_{xx}-B_{yy}-B_{zz}\right]
\end{align}
as the triplet and singlet contributions 
$(\textrm{d}\sigma_{s,m})/(\textrm{d}\Omega)$ which add up to 
$(\textrm{d}\sigma_0)/(\textrm{d}\Omega)$ given by Eq. \eqref{udcs}.  
Eq. \eqref{dcs} 
can also be expressed in the form
\begin{align}
\label{bilinears-np}
\frac{\textrm{d}\sigma}{\textrm{d}\Omega}=&\sum_{k_n,k_p,k}
\ \left(\left(P^{k_n}(n)\otimes P^{k_p}(p)\right)^k\cdot B^k(k_n,k_p)\right)\\
\label{analyse}
=&\frac{\textrm{d}\sigma_0}{\textrm{d}\Omega}\Biggl[1+\boldsymbol{P}(n)
\cdot\boldsymbol{A}(n)
+\boldsymbol{P}(p)\cdot\boldsymbol{A}(p)\nonumber\\&
\phantom{\frac{\textrm{d}\sigma_0}{\textrm{d}\Omega}\Biggl[}+\sum_{k=0}^2
\ \Bigl(\bigl(\boldsymbol{P}(n)\otimes\boldsymbol{P}(p)\bigr)^k
\cdot A^k\Bigr)\Biggr],
\end{align}
where the irreducible bilinear amplitudes in Eq. \eqref{bilinears-np}  
\begin{align}
\label{bilinear-np}
\hspace*{-\mathindent}&B^k_q(k_n,k_p)=\frac{3}{2}\left(\frac{k}{2\pi E}\right)^2
\ \frac{E_n E_p E_d}{p}\ (-1)^{k_n+k_p-k}\,[k_n]\,[k_p]\nonumber\\
&\times\sum_{s,s',\lambda,\lambda',\mu}
\ (-1)^\lambda\,[s]\,[s']\,[\lambda]\,[\lambda']
\,W(s'k1\lambda;s\lambda')\nonumber\\
&\times\left\{
\begin{matrix}
{\textstyle{\frac{1}{2}}}&{\textstyle{\frac{1}{2}}}&s\\[.2cm]
{\textstyle{\frac{1}{2}}}&{\textstyle{\frac{1}{2}}}&s'\\[.2cm]
k_n&k_p&k
\end{matrix}
\right\}
\left(T^\lambda(\mu,s)\otimes T^{\dagger^{\lambda'}}(\mu,s')\right)^k_q
\end{align}
are related to the spherical components of the analyzing powers in Eq. \eqref{analyse} through
\begin{align}
\label{A1qn}
&\frac{\textrm{d}\sigma_0}
{\textrm{d}\Omega}A^1_q(n)=B^1_q(1,0)\\[.15cm]
\label{A1qp}
&\frac{\textrm{d}\sigma_0}
{\textrm{d}\Omega}A^1_q(p)=B^1_q(0,1)\\[.15cm]
\label{Akqnp}
&\frac{\textrm{d}\sigma_0}
{\textrm{d}\Omega}A^k_q=B^k_q(1,1),
\end{align}
with $B^0_0(0,0)=(\textrm{d}\sigma_0)/(\textrm{d}\Omega)$, the unpolarized differential cross
section.
From Eq. \eqref{partial-wave}, we observe that the $T^\lambda_\nu(\mu,s)$
satisfy the symmetry property
\begin{align}
T^\lambda_\nu(\mu,s)=(-1)^{\lambda-\nu}\,T^\lambda_{-\nu}(-\mu,s).
\end{align}
This in turn implies through Eq. \eqref{bilinear-np} that the bilinear amplitudes 
satisfy the property
\begin{align}
\label{Bkq-symm}
B^k_q(k_n,k_p)=(-1)^{k-q}\,B^k_{-q}(k_n,k_p)
\end{align} 
so that 
\begin{align}
B^1_0(k_n,k_p)=0
\end{align}
and from Eqs. \eqref{A1qn} and \eqref{A1qp}, 
\begin{align}
A_z(n)=A_z(p)=0.
\end{align}
Using these results, we can now write the channel-spin differential cross sections
$(\text{d}\sigma_{s,m})/(\text{d}\Omega)$ in terms of the analyzing powers as
\begin{align}
\hspace*{-8pt}\frac{\text{d}\sigma_{1,1}}{\text{d}\Omega}&=
\frac{\text{d}\sigma_{1,-1}}{\text{d}\Omega}=
\frac{1}{4}\frac{\text{d}\sigma_0}{\text{d}\Omega}
\left[1-{\textstyle{\frac{1}{\sqrt{3}}}}A^0_0+
{\textstyle{\sqrt{\frac{2}{3}}}}A^2_0\right]\\[.2cm]
\hspace*{-8pt}\frac{\text{d}\sigma_{1,0}}{\text{d}\Omega}&=
\frac{1}{4}\frac{\text{d}\sigma_0}{\text{d}\Omega}
\left[1-{\textstyle{\frac{1}{\sqrt{3}}}}A^0_0-
2{\textstyle{\sqrt{\frac{2}{3}}}}A^2_0\right]\\[.2cm]
\hspace*{-8pt}\frac{\text{d}\sigma_{0,0}}{\text{d}\Omega}&=
\frac{1}{4}\frac{\text{d}\sigma_0}{\text{d}\Omega}
\left[1+\sqrt{3}A^0_0\right].  
\end{align}
The analyzing powers, moreover, are measurable readily in an experimental 
setup such as in \cite{muller}, where $\boldsymbol{P}(n)=\boldsymbol{P}(p)=
P\,\widehat{\boldsymbol{P}\,}$ with $P=50.5\%$ and $\widehat{\boldsymbol{P}\,}$ 
could be longitudinal as well as transverse.  With $\widehat{\boldsymbol{P}\,}$ 
chosen parallel and antiparallel to the beam, it is clear that
\begin{align}
\label{d_z}
\frac{\text{d}\sigma_{+z}}{\text{d}\Omega}+\frac{\text{d}\sigma_{-z}}
{\text{d}\Omega}=2\,\frac{\text{d}\sigma_{0}}{\text{d}\Omega}
\left[1+\frac{\ \,P^2}{\sqrt{3}}\bigl(\sqrt{2}\,A^2_0-A^0_0\bigr)\right]
\end{align}
which yields $\bigl(\sqrt{2}\,A^2_0-A^0_0\bigr)$.

Likewise, if $\widehat{\boldsymbol{P}\,}$ is chosen parallel and antiparallel to a 
direction (say $\hat{x}$ of $\hat{y}$) perpendicular to the beam, we have
\begin{align}
\label{d_x}
\frac{\text{d}\sigma_{+x}}{\text{d}\Omega}+\frac{\text{d}\sigma_{-x}}
{\text{d}\Omega}&=2\,\frac{\text{d}\sigma_{0}}{\text{d}\Omega}\Biggl[
1-\frac{\ \,P^2}{\sqrt{3}}\bigl(A^0_0+{\textstyle\frac{1}
{\sqrt{2}}}A^2_0\bigr)+A^2_2\Biggr]\\
\label{d_y}
\frac{\text{d}\sigma_{+y}}{\text{d}\Omega}+\frac{\text{d}\sigma_{-y}}
{\text{d}\Omega}&=2\,\frac{\text{d}\sigma_{0}}{\text{d}\Omega}\Biggl[
1-\frac{\ \,P^2}{\sqrt{3}}\bigl(A^0_0+{\textstyle\frac{1}
{\sqrt{2}}}A^2_0\bigr)-A^2_2\Biggr],
\end{align}
where we have made use of the fact that $A^2_2=A^2_{-2}$ from Eqs. \eqref{Akqnp} 
and \eqref{Bkq-symm}.
By adding Eqs. \eqref{d_x} and \eqref{d_y}, one readily obtains 
$\bigl(A^0_0+\frac{1}{\sqrt{2}}A^2_0\bigr)$.  Since $\bigl(\sqrt{2}\,A^2_0-A^0_0\bigr)$ 
is already known from Eq. \eqref{d_z}, one can determine $A^0_0$ as well as $A^2_0$.  
Hence all the $(\text{d}\sigma_{s,m})/(\text{d}\Omega)$ for $m=-s,\ldots,s$ 
and $s=0,1$ are determinable from experiment empirically at any given energy.  
We may perhaps add that these measurements need not have to be carried out at 
accuracies of $10^{-7}$ to determine $(\text{d}\sigma_{s,m})/(\text{d}\Omega)$.
Moreover, the discussion presented here aims to extend the program of
Breit and Rustgi \cite{breit-rustgi} to actual individual determinations 
of the three triplet and one singlet cross sections for the important
\cite{astro} fusion process at the differential level itself
rather that estimate the relative importance of the triplet
amplitude vis-a-vis the singlet amplitude.
It would, therefore, be desirable to extend the recent experiment \cite{muller} 
on $\vec{p}(\vec{n},\gamma)d$ to measure the observables \eqref{d_z} to \eqref{d_y}
in order to determine the differential cross sections
$(\text{d}\sigma_{s,m})/(\text{d}\Omega)$ individually and hence study 
the fusion reaction more incisively at any given energy.

It is perhaps not out of place to mention here that the highly
successful $NN$ potential models like Nijmegen1 and Nijmegen2
potentials \cite{Nijmegen}, the Argonne potential \cite{Argonne} and the
Bonn potential \cite{Bonn} which reproduce the elastic $NN$ scattering
data with a remarkable $\chi^2/{\text{datum}}\approx 1$, have
been subjected to an interesting study of the deuteron 
properties by Polls et al. \cite{polls}, who say,
{\textit{It is well known that the off-shell behavior of 
NN potentials cannot be pinned down by on-shell 
data...Moreover, within a given model, pinning down the on-shell point limits
the range of variation for the off-shell behavior.  The four modern
high-precision models considered in this study pin down the on-shell 
$T$-matrix as much as by all means possible since they fit the NN
scattering data with the perfect $\chi^2$/ datum $\approx$ 1.  Furthermore,
it may be reasonable to believe that the four models cover about the
range of diversity that there is to realistic physical models for
the NN interaction.  Based upon these premises, one may then conclude
that the off-shell uncertainties revealed in this study are what
we have to deal with, at the current status of theoretical nuclear
physics.  At this time, we do not know of any other objectively
verifiable aspects that could further reduce the off-shell uncertainties.
Tighter constraints for the off-shell behavior of NN may emerge in
the future...}}  
In this context, it is perhaps pertinent to point out that $np\to d\gamma$
involves necessarily off-shell $NN$ interactions.  Therefore
it would also be of interest to examine the fusion reaction 
with a view to throw light on the off-shell uncertainties.
Further work is in progress.    

\begin{acknowledgments}
One of us (GR) thanks Professor B. V. Srikantan for
much encouragement and Professor Ramanatha Cowsik for inviting him 
to the Indian Institute of Astrophysics, while the other (PND) 
thanks the Council of Scientific and Industrial Research (CSIR), India 
for support through the award of a Senior Research Fellowship.
\end{acknowledgments}


\begin{thebibliography}{99}
\bibitem{astro}
H. A. Bethe, Phys. Rev. {\textbf{55}}, 103 \& 434 (1939);\\
H. A. Bethe and C. Longmire, Phys. Rev. {\textbf{77}}, 647 (1950);\\
N. Austern and E. Rost, Phys. Rev. {\textbf{117}}, 1506 (1959);\\
R. V. Wagoner, W. A. Fowler and F. Hoyle, Astrophys. J. 
{\textbf{148}}, 3 (1967);\\
W. A. Fowler, G. R. Caughlam and B. A. Zimmerman, Ann. Rev. Astron. 
Astrophys. {\textbf{5}}, 525 (1967);\\
W. Rindler, Ann. Rev. Astron. Astrophys. {\textbf{11}}, 155 (1973);\\
D. N. Schramm and R. V. Wagoner, Ann. Rev. Nucl. Part. Sc. {\textbf{27}},
37 (1977);\\
R. A. Malaney and W. A. Fowler, {\textit{The Origin and Distribution of Elements}}, 
edited by G. J.~Mathews (World Scientific, Singapore, 1988);\\
C. F. Rolfs and W. S. Rodney, {\textit{Cauldrons in Cosmos}} (University of 
Chicago Press, Chicago, 1988);\\
E. W. Kolb and M. S. Turner, {\textit{The Early Universe}} (Addison-Wesley, 
Reading, 1990);\\
J. W. Chen and M. J. Savage, Phys. Rev. C {\textbf{60}}, 065205 (1999);\\
S. Burless, K. M. Nollett, J. W. Truran and M. S. Turner, Phys. Rev. Lett. 
{\textbf{82}}, 4176 (1999);\\
H. A. Bethe, Rev. Mod. Phys. {\textbf{71}}, S 6 (1999);\\
E. E. Salpeter, Rev. Mod. Phys. {\textbf{71}}, S 220 (1999);\\
K. Laganke and M. Wiescher, Rep. Prog. Phys. {\textbf{64}}, 1657 (2001). 
\bibitem{cs}
Y. Nagai {\textit{et al.,}} Phys. Rev. C {\textbf{56}}, 3173 (1997).
\bibitem{cs1}
A. E. Cox, S. A. R. Wynchank and C. H. Collie, Nucl. Phys. {\textbf{74}}, 497 (1965). 
\bibitem{mec}
D. O. Riska and G. E. Brown, Phys. Lett. B {\textbf{32}}, 193 (1972).
\bibitem{cs2}
I. Tuderic-Ghemo, Nucl. Phys. {\textbf{A92}}, 233 (1967);\\
M. Bosman {\textit{et al.,}} Phys. Lett. B {\textbf{82}}, 212 (1979).
\bibitem{ia}
H. Mathiot, Phys. Rep. {\textbf{173}}, 63 (1989)
\bibitem{inv}
H. Arenh\"ovel and M. Sanzone, {\textit{Photodisintegration of the Deuteron}} 
(Springer-Verlag, Berlin, 1991).
\bibitem{inv-expts}
F. D. Smith and F. D. Brooks, Nucl. Phys. {\textbf{A465}}, 429 (1987).
\bibitem{phot}
V. P. Likhachev {\textit{et al.,}} Nucl. Phys. {\textbf{A628}}, 597 (1998).
\bibitem{phot-rec}
E. C. Scheiber {\textit{et al.,}} Phys. Rev. C {\textbf{61}}, 061604(R) (2000).
\bibitem{angular}
J. Holt, K. Stephenson and J. R. Specht, Phys. Rev. Lett. {\textbf{50}}, 577 (1983). 
\bibitem{ang-neu}
P. Soderstrum and L. D. Knutson, Phys. Rev. C {\textbf{35}}, 1246 (1987).
\bibitem{recent1}
M. S. Smith, L. H. Kawano and R. A. Malaney, Astrophys. J. 
Suppl. Ser. {\textbf{85}}, 219 (1993).
\bibitem{recent2}
T. Sato, M. Niwa and H. Ohtsubo, in {\textit{Proceedings of the International Symposium on 
Weak and Electromagnetic Interactions in Nuclei}}, edited by H. Ejiri, 
T. Kashimoto and T. Sato (World Scientific, Singapore, 1995) p 488.
\bibitem{old}
G. R. Bishop {\textit{et al.,}} Phys. Rev. {\textbf{80}}, 211 (1950).
\bibitem{breit-rustgi}
G. Breit and M. L. Rustgi, Nucl. Phys. {\textbf{A161}}, 337 (1971).
\bibitem{J-W}
J. -W. Chen, G. Rupak and M. J. Savage, Phys. Lett. B {\textbf{464}}, 1 (1999).
\bibitem{bazhanov}
A. N. Bazhanov {\textit{et al.,}} Phys. Lett. B {\textbf{289}}, 17 (1992).
\bibitem{muller}
T. M. M\"uller, D. Dubbers, D. Hautle and O. Zimmer, 
in {\textit{Proceedings of the International Workshop on Particle Physics with 
Slow Neutrons}}, ILL, Grenoble, France, 22-24 Oct 1998;\\
T. M. M\"uller {\textit{et al.,}} Nucl. Instr. Meth. {\textbf{A440}}, 736 (2000).
\bibitem{wein}
S. Weinberg, Phys. Lett. B {\textbf{251}}, 288 (1990);
\ Nucl. Phys. {\textbf{B363}}, 3 (1991);\ Phys. Lett. B {\textbf{295}}, 114 (1992).
\bibitem{park}
T. -S. Park, K. Kubodera, D. -D. Min and M. Rho, Phys. Lett. B {\textbf{472}}, 232 (2000).
\bibitem{bug}
D. V. Bugg, Ann. Rev. Nucl. Part. Sc. {\textbf{35}}, 295 (1985).  
\bibitem{NN}
W. S. Wilburn {\textit{et al.,}} Phys. Rev. Lett. {\textbf{71}}, 1982 (1993);\\
W. S. Wilburn {\textit{et al.,}} Phys. Rev. C {\textbf{52}}, 235 (1995);\\
J. Bro\v z {\textit{et al.,}} Z. Phys. A {\textbf{354}}, 401 (1996);\\
J. Bro\v z {\textit{et al.,}} Z. Phys. A {\textbf{359}}, 23 (1997);\\
R. W. Raichle {\textit{et al.,}} Phys. Rev. Lett. {\textbf{83}}, 2711 (1999);\\
J. R. Walston {\textit{et al.,}} Phys. Rev. C {\textbf{63}}, 014004 (2000).
\bibitem{pp}
H. O. Meyer {\textit{et al.,}} Phys. Rev. Lett. {\textbf{81}}, 3096 (1998);\\
P. Th\"orngren-Engblom {\textit{et al.,}} Contribution to the Conference 
``Mesons and Light Nuclei'', 1998, Prague-Pruhonice, Czech Republic, nucl-ex/9810013;\\
H. O. Meyer {\textit{et al.,}} Phys. Rev. Lett. {\textbf{83}}, 5439 (1999);\\
P. Th\"orngren-Engblom {\textit{et al.,}} Nucl. Phys. {\textbf{A663\&664}}, 447 (2000);\\
H. O. Meyer {\textit{et al.,}} Phys. Rev. C {\textbf{63}}, 064002 (2001).
\bibitem{cjr}
C. Cohen, D. L. Judd and R. J. Riddell Jr., Phys. Rev. 
{\textbf{119}}, 384 (1960).
\bibitem{friar}
J. L. Friar, B. F. Gibson, H. C. Jean and 
G. L. Payne, Phys. Rev. Lett. {\textbf{66}}, 1827 (1991);\\
M. Viviani, R. Schiavila and A. Kievsky, Phys. Rev. C {\textbf{54}}, 534 (1996);\\
M. Viviani, A. Kievsky, L. E. Marcucci, S. Rosatti and R. Schiavila, 
Phys. Rev. C {\textbf{61}}, 064001 (2000).
\bibitem{grpndsp}
G. Ramachandran, P. N. Deepak and S. Prasanna Kumar, 
J. Phys. G: Nucl. Part. Phys. {\textbf{25}}, L155 (1999).
\bibitem{grpnd}
G. Ramachandran and P. N. Deepak, Nucl. Phys. {\textbf{A695}}, 177 (2001). 
\bibitem{mad}
G. R. Satchler {\textit{et al.,}} in {\textit{Proceedings of the 3rd International 
Symposium on 
Polarization Phenomena in Nuclear Reactions}}, edited by H.H. Barschall and 
W. Haeberli, (University of Wisconsin Press, Madison, 1970).
\bibitem{rose}
M. E. Rose, {\textit{Elementary Theory of Angular Momentum}}, (John Wiley, New York, 1957).
\bibitem{gr-msv}
G. Ramachandran and M. S. Vidya, Phys. Rev. C {\textbf{56}}, R12 (1997);\\
G. Ramachandran, M. S. Vidya and M. M. Prakash, Phys. Rev. C 
{\textbf{56}}, 2882 (1997).
\bibitem{rus}
D. A. Varshalovich, A. N. Moskalev and V. K. Khersonskii,
{\textit{Quantum Theory of Angular Momentum}} (World Scientific, Singapore, 1988).
\bibitem{wolf}
L. Wolfenstein and J. Ashkin, Phys. Rev. {\textbf{85}}, 947 (1952);\\
L. Wolfenstein, Phys. Rev. {\textbf{96}}, 1654 (1954);\\
C. Lechanoine-Leluc and F. Lehar, Rev. Mod. Phys. {\textbf{65}}, 47 (1993).
\bibitem{prc}
G. Ramachandran and P. N. Deepak, Phys. Rev. C {\textbf{63}}, 051001(R) (2001);\\
P. N. Deepak and G. Ramachandran, Phys. Rev. C {\textbf{65}}, 027601 (2002).
\bibitem{Nijmegen}
V. G. J. Stoks, R. A. M. Klomp, C. P. F. Terheggen and J. J. de Swart, Phys.
Rev. C {\textbf{49}}, 2950 (1994).
\bibitem{Argonne}
R. B. Wiringa, V. G. J. Stoks and R. Schiavilla, Phys. Rev. C {\textbf{51}},
38 (1995).
\bibitem{Bonn}
R. Machleidt, F. Sammarruca and Y. Song, Phys. Rev. C 
{\textbf{53}}, 1483 (1996);\\
R. Machleidt, Phys. Rev. C {\textbf{63}}, 024001 (2001).
\bibitem{polls}
A. Polls, H. M\"uther, R. Machleidt and M. Hjorth-Jensen, Phys. Lett. B {\textbf{432}},
1 (1998).
\end{thebibliography}
\end{document}